\documentclass[pre,aps]{revtex4}%
\usepackage{amsmath}
\usepackage{amsfonts}
\usepackage{amssymb}
\usepackage{graphicx}%
\providecommand{\U}[1]{\protect\rule{.1in}{.1in}}
\begin{document}
\title{The hidden phase of Fock states; quantum non-local effects}
\author{F.\ Lalo\"{e}, LKB, Dept. de Physique de l'ENS, 24 rue Lhomond, 75005 Paris, France}
\date{\today}

\begin{abstract}
We revisit the question of how a definite phase between Bose-Einstein
condensates can spontaneously appear under the effect of measurements.

We first consider a system that is the juxtaposition of two subsystems in Fock
states with high populations, and assume that successive individual position
measurements are performed. Initially, the relative phase is totally
undefined, and no interference effect takes place in the first position
measurement.\ But, while successive measurements are accumulated, the relative
phase becomes better and better defined, and a clear interference pattern
emerges. It turns out that all observed results can be interpreted in terms of
a pre-existing, but totally unknown, relative phase, which remains exactly
constant during the experiment.

We then generalize the results to more condensates.\ We also consider other
initial quantum states than pure Fock states, and distinguish between
intrinsic phase of a quantum state and phase induced by
measurements.\ Finally, we examine the case of multiple condensates of spin
states.\ We discuss a curious quantum effect, where the measurement of the
spin angular momentum of a small number of particles can induce a big angular
momentum in a much larger assembly of particles, even at an arbitrary
distance.\ This spin observable can be macroscopic, analogous to the pointer
of a measurement apparatus, which illustrates the non-locality of standard
quantum mechanics with particular clarity. The effect can be described as the
teleportation at arbitrary distances of the continuous classical result of a
local experiment. The EPR argument, transposed to this case, takes a
particularly convincing form since it does not involve incompatible
measurements and deals only with macroscopic variables.

\end{abstract}
\maketitle

PACS 03.65. Ta and Ud ; 03.75.Gg

\bigskip

It is well known in quantum mechanics that one cannot construct a quantum
state where the number of particles and the phase are both arbitrarily well
defined; they are actually incompatible observables, related by a Heisenberg
type uncertainty relation, as position and momentum of a single particle.\ For
a general discussion of the phase operator in quantum mechanics, see for
instance refs. \cite{D, CN,AD}.\ An usual illustration of the relation between
phase and particle number is given by the so called \textquotedblleft coherent
states\textquotedblright\ or \textquotedblleft Schr\"{o}dinger semi-classical
states\textquotedblright, where fluctuations of the number of particles are
used in order to define a phase.\ These states are often discussed in the
context of electromagnetism \cite{G}, but they also apply to Bose-Einstein
condensates, on which we focus the interest here - more precisely, multiple
condensates and their relative phase.

For instance we consider a physical system that is the juxtaposition of two
systems in Fock states (number states), described by the state vector:%
\begin{equation}
\mid\Phi_{0}>~=~\mid N_{a}:\varphi_{a}\ ;~N_{b}:\varphi_{b}> \label{0}%
\end{equation}
where $N_{a}$ particles are condensed in the same single particle quantum
state $\varphi_{a}$, and $N_{b}$ particles in the same quantum state
$\varphi_{b}$. These two states may correspond to plane waves with momenta
$\mathbf{k}_{a}$ and $\mathbf{k}_{b}$, a case in which one can expect the
occurrence of an interference pattern with spatial frequency $\mathbf{k}%
_{a}-\mathbf{k}_{b}$.\ Nevertheless, the position of this pattern depends of
the relative phase of the waves, which is completely unknown in a state such
as (\ref{0}).\ One could then wonder whether an interference effect is
observable at all under these conditions.

The question was studied theoretically by several authors \cite{JH, WCW, CGNZ,
CD, M-1}; see also refs. \cite{M-2, CCT, CH, PS}.\ The result is that an
interference pattern is in fact observed, with a phase that spontaneously
emerges from the quantum measurement process itself.\ The detection of the
first particle occurs at a completely random position, but this position
provides a first information on the value of the phase; the second particle is
then detected at a position that is correlated with the first, and this
measurements makes the information on the phase more precise; as more and more
detection events are accumulated, the phase becomes better and better
defined.\ In practice, a relatively small number of detections is sufficient
to determine the relative phase with good accuracy.\ Nevertheless, before the
first measurement, according to standard quantum mechanics there is no way
whatsoever to predict the value that this phase will spontaneously choose. In
other words, repeating the same experiment from the same initial conditions
(\ref{0}) will lead to another, completely independent, value of the relative
phase. The spontaneous appearance of a relative phase under the effect of
measurement is related to the notion of spontaneous symmetry breaking of the
conservation of the number of particles, and to non-zero average values of the
field operator in superfluid systems \cite{A-1, LS}.\ It also relates to
Anderson's intriguing question, as quoted by Leggett \cite{Leggett}:
\textquotedblleft Do superfluids which have never seen each other possess a
definite relative phase?\textquotedblright\ \cite{A-2, A-3}.

In this article, we revisit this question by using a slightly different method
from that of refs. \cite{JH, WCW, CGNZ, CD, M-1, M-2, CCT, PS}; we propose a
straightforward analytical calculation that does not require the introduction
of any (incoherent) fluctuation of the particle numbers, as in \cite{CGNZ} and
\cite{CD} for instance, but deals directly with pure Fock states; it
reproduces the stochastic simulations of \cite{JH} with only one approximation
(total number of particles much larger than the number of measurements).\ In
\S \ref{spontaneous} we discuss how the process of measurement, applied to a
pair of systems that are initially in Fock states, therefore with no well
defined phase, can make such a phase appear spontaneously ; in
\S \ref{general} we apply our calculations to more general cases and discuss
how an intrinsic phase, contained in the initial state, can combine with a
spontaneous phase induced by measurements; in \S \ref{spin}, we discuss
multiple condensates in the context of spin states and show in particular how
the measurement of a small number of spins may induce an angular momentum for
a much larger assembly of spins, a curious non-local quantum effect.

\section{Spontaneous phase induced by measurements on Fock states}

\label{spontaneous}

We first recall a general result of quantum mechanics.\ We consider a system
in a quantum state $\mid\Psi_{0}>$, and a series of observables:%
\begin{equation}%
\begin{array}
[c]{lll}%
A & \text{with~eigenvalues~}a_{i}\text{;} & P(a_{i})\text{ is the projector
onto the corresponding eigenstate(s)}\\
B & \text{with~eigenvalues~}b_{j}\text{;} & P(b_{j})\text{ is the projector
onto the corresponding eigenstate(s) }\\
C & \text{etc.} & \text{etc.}%
\end{array}
\label{1}%
\end{equation}
If we assume that all these observables commute with each other, then the
probability of observing in a combined measurement the results $a_{i}$,
$b_{j}$, $c_{k}$, etc. can simply be written:%
\begin{equation}
<\Phi_{0}\mid P(a_{i})~P(b_{j})~P(c_{k})..\mid\Phi_{0}> \label{2}%
\end{equation}
(the usual intrication of projectors in the so called Wigner formula is not
necessary here because the operators commute). Whether or not the wave packet
reduction postulate is applied does no matter; in other words, all
measurements can be made simultaneously, or one after the other, and the order
of measurements is irrelevant. What is assumed, nevertheless, is that the
sequence of measurements covers a time that is negligible in comparison with
the time constants associated with the intrinsic evolution of the physical
system; this is why no evolution operator has to be introduced in the
formula.\ We now apply (\ref{2}) to position measurements inside the overlap
regions of two or more condensates.

\subsection{A simple case: two highly populated states}

\label{simple}

We begin with a simple case, where the system is just a juxtaposition of two
Bose-Einstein condensates in plane waves:%
\begin{equation}
\mid\Phi_{0}>~=~\mid N_{a}:\mathbf{k}_{a}\ ;~N_{b}:\mathbf{k}_{b}> \label{3}%
\end{equation}
Here $\mathbf{k}_{a}$ and $\mathbf{k}_{b}$ are single particle states of well
defined momentum normalized in a box with periodic boundary conditions; we
assume that $N_{a}$ and $N_{b}$ are large numbers.$\ $The probability for
detecting a particle at point $\mathbf{r}$ corresponds to the following
operator, similar to the projectors $P(a_{i})$ introduced above:%
\begin{equation}
\Psi^{\dagger}(\mathbf{r})~\Psi(\mathbf{r})=\sum_{i=1}^{N}~\mid i:\mathbf{r}%
><i:\mathbf{r\mid} \label{4}%
\end{equation}
Here $\Psi(\mathbf{r})$ is the usual field operator, and $\mid i:\mathbf{r}>$
is the one particle state corresponding to a perfect localization of particle
$i$ at point $\mathbf{r}$; $N=N_{a}+N_{b}$ is the total number of
particles.\ To make the operator $\Psi^{\dagger}(\mathbf{r})~\Psi(\mathbf{r})$
really similar to a projector with eigenvalues $0$ and $1$, one has to
integrate it over some small $\mathbf{r}$ domain $\Delta_{\mathbf{r}}$
centered around $\mathbf{r}$; assuming that $\Delta_{\mathbf{r}}$ is
sufficiently small ensures that the probability of finding two particles or
more in $\Delta_{\mathbf{r}}$ is negligible.\ For the sake of simplicity we do
not write these integrations explicitly; in other words we write probability
densities instead of probabilities, but it would be easy to come back to real
probabilities by multiplying by an appropriate power of $\Delta_{\mathbf{r}}%
$.\ The field operator can be expanded onto the annihilation operators
$a_{\mathbf{k}}$ of momentum states according to:%
\begin{equation}
\Psi(\mathbf{r})\sim~\sum_{\mathbf{k}}~e^{i\mathbf{k}\cdot\mathbf{r}%
}~a_{\mathbf{k}}~ \label{5}%
\end{equation}
where, as usual, the sum over $\mathbf{k}$ ranges over the values allowed by
periodic boundary conditions in a box.

We now calculate the probability for observing one particle at position
$\mathbf{r}_{1}$, another at position $\mathbf{r}_{2}$, still another at
position $\mathbf{r}_{3}$, .. one at position $\mathbf{r}_{P}$.\ We consider
that all positions $\mathbf{r}_{1}$, $\mathbf{r}_{2}$, ..are different, so
that all operators $\Psi$ and $\Psi^{\dagger}$ commute (more precisely, we
assume that the small integration domains $\Delta_{\mathbf{r}_{1}}$,
$\Delta_{\mathbf{r}_{21}}$, .. do not overlap).\ This allows us to write this
probability as:%
\begin{equation}
<\Phi_{0}\mid\Psi^{\dagger}(\mathbf{r}_{1})\Psi^{\dagger}(\mathbf{r}_{2}%
)\Psi^{\dagger}(\mathbf{r}_{3})...\Psi(\mathbf{r}_{3})\Psi(\mathbf{r}_{2}%
)\Psi(\mathbf{r}_{1})\mid\Phi_{0}> \label{6}%
\end{equation}
or, if (\ref{5}) is used:%
\begin{equation}%
\begin{array}
[c]{l}%
\sim\sum_{\mathbf{k}_{1}}\sum_{\mathbf{k}_{1}^{^{\prime}}}e^{i(\mathbf{k}%
_{1}-\mathbf{k}_{1}^{^{\prime}})\cdot\mathbf{r}_{1}}\sum_{\mathbf{k}_{2}}%
\sum_{\mathbf{k}_{2}^{^{\prime}}}~e^{i(\mathbf{k}_{2}-\mathbf{k}_{2}%
^{^{\prime}})\cdot\mathbf{r}_{2}}\times\\
<\Phi_{0}\mid a_{\mathbf{k}_{1}^{^{\prime}}}^{\dagger}~a_{\mathbf{k}%
_{2}^{^{\prime}}}^{\dagger}.......a_{\mathbf{k}_{P}^{^{\prime}}}^{\dagger
}~a_{\mathbf{k}_{P}}......a_{\mathbf{k}_{2}}a_{\mathbf{k}_{1}}\mid\Phi_{0}>
\end{array}
\label{7}%
\end{equation}
Acting on $\mid\Phi_{0}>$ given by (\ref{3}), the annihilation operators $a$'s
in the second line of this formula give zero unless the values of the
$\mathbf{k}$ and $\mathbf{k}^{^{\prime}}$ momenta are all equal to
$\mathbf{k}_{a}$ or $\mathbf{k}_{b}$. In addition, the series of $\mathbf{k}%
$'s must contain the same number of $\mathbf{k}_{a}$'s (and $\mathbf{k}_{b}%
$'s) as the series of $\mathbf{k}^{^{\prime}}$'s; otherwise, the product of
$a$ and $a\dagger$ operators creates a ket orthogonal to $\mid\Phi_{0}>$, and
the second line vanishes. When this condition is met, each $a_{\mathbf{k}%
_{a,b}}$ or $a_{\mathbf{k}_{a,b}}^{\dagger}$ introduces a factor
$\sqrt{N_{a,b}\pm n}$, where $n$ is some integer ranging from $0$ and $P$ (the
number of measurements).

At this point we make our only approximation: we assume that $P\,\ $is much
smaller than $N_{a}$ and $N_{b}$, so that we can replace all these factors by
$\sqrt{N_{a,b}}$. Now, in the first line of (\ref{7}), consider for instance
the two first summations over $\mathbf{k}_{1}$ and $\mathbf{k}_{1}^{^{\prime}%
}$; if one of these vectors is chosen as $\mathbf{k}_{a}$, the other as
$\mathbf{k}_{b}$, one obtains a contribution $\sqrt{N_{a}N_{b}}e^{\pm i\left(
\mathbf{k}_{a}-\mathbf{k}_{b}\right)  \cdot\mathbf{r}_{1}}$; if both are equal
to the same value, $\mathbf{k}_{a}$ or $\mathbf{k}_{b}$, one gets a
contribution $N_{a}+N_{b}=N$. The same result is obviously also valid for all
other sums.\ Therefore, if we introduce the notation:%
\begin{equation}
\left\{
\begin{array}
[c]{l}%
F_{0}(\mathbf{r)}~=~N_{a}+N_{b}=N\\
F_{\pm1}(\mathbf{r})~=\sqrt{N_{a}N_{b}}e^{\pm i(\mathbf{k}_{a}-\mathbf{k}%
_{b})\cdot\mathbf{r}}=xN~e^{\pm i(\mathbf{k}_{a}-\mathbf{k}_{b})\cdot
\mathbf{r}}/2
\end{array}
\right.  \label{8}%
\end{equation}
where $x$ is the contrast ratio $x$ ($0\leq x\leq1$) defined by:%
\begin{equation}
N=N_{a}+N_{b}~~~~~;~~~~~x=\frac{2\sqrt{N_{a}N_{b}}}{N} \label{9}%
\end{equation}
we can rewrite (\ref{7}) as:%
\begin{equation}
\sim\sum_{\left\{  \sum~q_{i}=0\right\}  }F_{q_{1}}(\mathbf{r}_{1})~F_{q_{2}%
}(\mathbf{r}_{2})..............F_{q_{P}}(\mathbf{r}_{P}) \label{10}%
\end{equation}
In this expression, the sum contains all possible values $0$, $\pm1$ of
$q_{1}$, $q_{2}$, etc.., with the only condition that their sum be zero. Now,
since the integral:%
\begin{equation}
\int_{0}^{2\pi}\frac{d\Phi}{2\pi}~e^{iq\Phi} \label{11}%
\end{equation}
(where $q$ is a positive or negative integer) vanishes except if $q=0$, we can
release this condition by rewriting (\ref{10}) as:%
\begin{equation}
\sim\int_{0}^{2\pi}\frac{d\Phi}{2\pi}~%
%TCIMACRO{\dprod \limits_{i=1}^{P}}%
%BeginExpansion
{\displaystyle\prod\limits_{i=1}^{P}}
%EndExpansion
\left\{  F_{0}(\mathbf{r}_{i})+e^{i\Phi}F_{+1}(\mathbf{r}_{i})+e^{-i\Phi
}F_{-1}(\mathbf{r}_{i})\right\}  \label{12}%
\end{equation}
Finally, inserting (\ref{8}) into this result provides the probability in the
form:%
\begin{equation}
\sim\int_{0}^{2\pi}\frac{d\Phi}{2\pi}~%
%TCIMACRO{\dprod \limits_{i=1}^{P}}%
%BeginExpansion
{\displaystyle\prod\limits_{i=1}^{P}}
%EndExpansion
~\left\{  1+x\ \cos\left[  (\mathbf{k}_{a}-\mathbf{k}_{b})\cdot\mathbf{r}%
_{i}+\Phi\right]  \right\}  \label{13}%
\end{equation}
This result is simple to interpret: the $\mathbf{r}_{i}$ dependence of any
sequence of events is exactly the same as for a system with a well defined,
but completely unknown, initial phase $\Phi$.\ For each value of this phase,
the $\mathbf{r}_{i}$ dependence of the probability factorizes and the
detection events are independent; it gives exactly the usual interference
pattern between two waves with wave vectors $\mathbf{k}_{a}$ and
$\mathbf{k}_{b}$.\ But the summation of $\Phi$ between $0$ and $2\pi$ destroys
the factorization and correlates the events.

Let us assume for the sake of simplicity that the contrast ratio $x$ is equal
to $1$.\ When the first detection localizes a particle at point $\mathbf{R}%
_{1}$, some information on $\Phi$ is immediately obtained, if only because the
value $\Phi=(\mathbf{k}_{a}-\mathbf{k}_{b})\cdot\mathbf{R}_{1}+\pi$ becomes
incompatible with the first detection.\ After this measurement, the
probability of a second detection at point $\mathbf{r}_{2}$ is proportional
to:%
\begin{equation}
\int_{0}^{2\pi}\frac{d\Phi}{2\pi}\left(  1+\cos\left[  (\mathbf{k}%
_{a}-\mathbf{k}_{b})\cdot\mathbf{R}_{1}+\Phi\right]  \right)  \left\{
1+\cos\left[  (\mathbf{k}_{a}-\mathbf{k}_{b})\cdot\mathbf{r}_{2}+\Phi\right]
\right\}  \label{14}%
\end{equation}
where the distribution of the values of $\Phi$ is no longer uniform: it is now
given by the $\mathbf{R}_{1}$ dependent sinusoidal function between the first
brackets inside the integral. If two measurements are initially performed with
localizations at points $\mathbf{R}_{1}$ and $\mathbf{R}_{2}$, the $\Phi$
distribution is then given by the product of two sinusoidal functions, which
is more peaked than the previous distribution, etc.. More and more precise
information on the value of $\Phi$ is progressively accumulated while more and
more measurements are performed.\ At some point, the phase is practically
determined and the detection events become quasi independent. As pointed out
by M\o lmer \cite{M-3}, this is because the two condensates tend to fuse into
a single condensate under the effect of successive interference
measurements.\ See also ref.\ \cite{HB} for discussion of the evolution of the
system towards a coherent state, which requires a number of detections
comparable to the total number of atoms.\ For more details about the evolution
of the $\Phi$ distribution function, see for instance \cite{CGNZ, CD, CCT}.

It is interesting to note that the phase $\Phi$ plays a role which appears,
mathematically, very similar to that of the so called \textquotedblleft
additional variables\textquotedblright\ (or \textquotedblleft hidden
variables\textquotedblright) sometimes introduced to interpret the results of
quantum mechanics - see for instance \cite{Bell-0} and \cite{FL-1} for a
review.\ If standard quantum mechanics is completed with an additional phase
variable, one can consider that the measurement process \textit{reveals} a
pre-existing value of $\Phi$, instead of \textit{creating} the relative phase
between the two condensates as in the standard interpretation of quantum
mechanics. In our calculation, the motivation for introducing $\Phi$ was not a
fundamental re-interpretation of quantum mechanics; it was just a convenient
way to sum the many terms in (\ref{7}) while maintaining the sum rule over the
series of values of $\mathbf{k}$ and $\mathbf{k}^{^{\prime}}$.\ The precise
role of the $\Phi$ integral is to ensure that the successive creation and
annihilation operators bring back the system to the same initial Fock state,
in other words to enforce a number conservation rule.\ Each term in the sum
can be seen as arising from one possible path over intermediates states in
which the values of the occupation numbers vary in a given way before coming
bask to their initial values.\ It is interesting to remark that the
interference of all these paths should result in a simple phase integral,
leading naturally to the introduction of an additional variable $\Phi$. We
recover the usual relation between conjugate variables, where a sum over one
variable with equal weights ensures a strict conservation of the other; the
phase appears here as related to fluctuations of the number of particles in
intermediate states (instead of in the initial state as usual).

\subsection{Three highly populated states}

\label{three}

We now generalize the preceding discussion to three states by assuming that
the initial state of the system is given by:%
\begin{equation}
\mid\Phi_{0}>~=~\mid N_{a}:\mathbf{k}_{a};~N_{b}:\mathbf{k}_{b};~N_{c}%
:\mathbf{k}_{c}> \label{15}%
\end{equation}
and evaluate the probability of particle position measurements at different
points $\mathbf{r}_{1}$, $\mathbf{r}_{2}$, .. etc. by using (\ref{7})
again.\ The calculation remains very similar to that of the preceding section.
The only difference is that the $\mathbf{k}$ and $\mathbf{k}^{^{\prime}}$ can
now take three values, $\mathbf{k}_{a}$, $\mathbf{k}_{b}$ and $\mathbf{k}_{c}%
$, so that in the second line of (\ref{7}) it is no longer sufficient to
ensure that $N_{a}$ returns to the same value; the value of $N_{b}$ (or
$N_{c}$) must also be controlled.\ This can be obtained by introducing two
relative phases, $\Phi$ and $\Phi^{^{\prime}}$. Another difference is that the
sums in the first line of (\ref{7}) now include three terms that contain the
same $\mathbf{k}$'s, as well as six others that contain different $\mathbf{k}%
$'s and introduce cosines.\ Therefore, if we define:%
\begin{equation}
N=N_{a}+N_{b}+N_{c}\text{ \ \ \ \ \ \ ; \ \ \ \ \ \ \ }x_{ab}^{^{\prime}%
}=\frac{3\sqrt{N_{a}N_{b}}}{N} \label{16}%
\end{equation}
we obtain the quantity:%
\begin{equation}
N\left\{  1+\frac{2x_{ab}}{3}\cos\left[  (\mathbf{k}_{a}-\mathbf{k}_{b}%
)\cdot\mathbf{r}_{i}+\Phi\right]  +\frac{2x_{bc}}{3}\cos\left[  (\mathbf{k}%
_{b}-\mathbf{k}_{c})\cdot\mathbf{r}_{i}+\Phi^{^{\prime}}\right]
+\frac{2x_{ca}}{3}\cos\left[  (\mathbf{k}_{c}-\mathbf{k}_{a})\cdot
\mathbf{r}_{i}-\Phi-\Phi^{^{\prime}}\right]  \right\}  \label{17}%
\end{equation}
and the probability is proportional to:%
\begin{equation}
\sim\int_{0}^{2\pi}\frac{d\Phi}{2\pi}\int_{0}^{2\pi}\frac{d\Phi^{^{\prime}}%
}{2\pi}%
%TCIMACRO{\dprod \limits_{i=1}^{P}}%
%BeginExpansion
{\displaystyle\prod\limits_{i=1}^{P}}
%EndExpansion
\left\{  \frac{3}{2}+\cos\left[  (\mathbf{k}_{a}-\mathbf{k}_{b})\cdot
\mathbf{r}_{i}+\Phi\right]  +\cos\left[  (\mathbf{k}_{b}-\mathbf{k}_{c}%
)\cdot\mathbf{r}_{i}+\Phi^{^{\prime}}\right]  +\cos\left[  (\mathbf{k}%
_{c}-\mathbf{k}_{a})\cdot\mathbf{r}_{i}+-\Phi-\Phi^{^{\prime}}\right]
\right\}  \label{19}%
\end{equation}
The situation is therefore a direct generalization of that studied in the
preceding section. Initially, the two phases $\Phi$ and $\Phi^{^{\prime}}$ are
completely undetermined but, as more and more position measurements are
recorded, the relative phases become better and better determined. We can
generalize to $K$ systems in highly populated Fock states, for which position
measurements progressively determine the value of $K-1$ phase
differences.\ The phases have a property of transitivity \cite{Leggett-phase,
DB}: measuring the $a-b$ phase difference and the $b-c$ phase difference
provides by sum the knowledge of the $a-c$ phase difference.

Note however that, in all our calculations, we have ignored the intrinsic
evolution of the system; as mentioned above, we have assumed that all
measurements are made in a time that is sufficiently short to justify this
approximation.\ Over longer periods of time, the inherent evolution of the
condensates takes place and introduces phase dynamics.\ Any term in the
Hamiltonian containing occupation numbers will tend to compete with the
definition of phase introduced by the measurements \cite{LS, ZSL}; for
instance, the mean-field interactions cause a quantum spreading of the phase
\cite{KS}.\ These effects are not included in the present article.

\section{Intrinsic and induced phase}

\label{general}

Until now, we have limited our study to initial states of the system that are
products of Fock states, where the initial occupation numbers are perfectly
well defined.\ We will now allow these numbers to fluctuate, beginning with
the simple case where their difference fluctuates but their sum remains constant.

\subsection{Constant total number of particles}

We now assume that:%
\begin{equation}
\mid\Phi_{0}>~=~%
%TCIMACRO{\dsum \limits_{Q=0}^{Q_{\max.}}}%
%BeginExpansion
{\displaystyle\sum\limits_{Q=0}^{Q_{\max.}}}
%EndExpansion
~x_{Q}~\mid N_{a}+Q:\mathbf{k}_{a};~N_{b}-Q:\mathbf{k}_{b}~> \label{20}%
\end{equation}
where the difference between the population numbers varies between
$N_{a}-N_{b}$ and $N_{a}-N_{b}+2Q$.; we will assume that $Q\ll N_{a},N_{b}$.
\ The calculation remains similar to that of \S \ref{spontaneous}, the main
difference being that a double sum:%
\begin{equation}
\sum_{Q,Q^{^{\prime}}}~x_{Q}^{\ast}x_{Q^{^{\prime}}} \label{21}%
\end{equation}
is added in the calculation of the probability.\ In each term of this sum,
instead of coming back to the same value of the difference $N_{a}-N_{b}$ under
the effect of the sequence of $a$ and $a\dagger$ operators, a change
$Q-Q^{^{\prime}}$ has to be accumulated in order to obtain a non vanishing
contribution.\ Equation (\ref{10}) is therefore now replaced by:%
\begin{equation}%
%TCIMACRO{\dsum \limits_{Q,Q^{^{\prime}}}}%
%BeginExpansion
{\displaystyle\sum\limits_{Q,Q^{^{\prime}}}}
%EndExpansion
~x_{Q}^{\ast}~x_{Q^{^{\prime}}}\sum_{\left\{  \sum q_{i}=Q-Q^{^{\prime}%
}\right\}  }F_{q_{1}}(\mathbf{r}_{1})~F_{q_{2}}(\mathbf{r}_{2}).......F_{q_{P}%
}(\mathbf{r}_{P}) \label{22}%
\end{equation}
or:%
\begin{equation}%
%TCIMACRO{\dsum \limits_{Q,Q^{^{\prime}}}}%
%BeginExpansion
{\displaystyle\sum\limits_{Q,Q^{^{\prime}}}}
%EndExpansion
~x_{Q}^{\ast}~x_{Q^{^{\prime}}}~\int\frac{d\Phi}{2\pi}e^{i(Q-Q^{^{\prime}%
})\Phi}%
%TCIMACRO{\dprod \limits_{i=1}^{P}}%
%BeginExpansion
{\displaystyle\prod\limits_{i=1}^{P}}
%EndExpansion
\left\{  F_{0}(\mathbf{r}_{i})+e^{i\Phi}F_{+1}(\mathbf{r}_{i})+e^{-i\Phi
}F_{-1}(\mathbf{r}_{i})\right\}  \label{23}%
\end{equation}
This naturally leads to the introduction of the real positive function:%
\begin{equation}
G(\Phi)=c\left\vert
%TCIMACRO{\dsum \limits_{Q}}%
%BeginExpansion
{\displaystyle\sum\limits_{Q}}
%EndExpansion
~x_{Q}~e^{-iQ\Phi}\right\vert ^{2} \label{24}%
\end{equation}
where $c$ is a normalization factor ensuring that the integral of $G(\Phi)$
between $0$ and $2\pi$ is $1$.\ We can now write the probability as:%
\begin{equation}
\sim\int_{0}^{2\pi}\frac{d\Phi}{2\pi}G(\Phi)%
%TCIMACRO{\dprod \limits_{i=1}^{P}}%
%BeginExpansion
{\displaystyle\prod\limits_{i=1}^{P}}
%EndExpansion
~\left\{  1+x~\cos\left[  (\mathbf{k}_{a}-\mathbf{k}_{b})\cdot\mathbf{r}%
_{i}+\Phi\right]  \right\}  \label{25}%
\end{equation}
which is straightforward to interpret: instead of being completely unknown
between $0$ and $2\pi$, the initial phase $\Phi$ before the first measurements
now has a distribution function $G(\Phi)$.\ This function is given by a
Fourier transform of the probability amplitudes associated with the initial
variations of the difference $N_{a}-N_{b}$ contained in (\ref{20}). When
successive position measurements begin, they all contribute to a further
determination of the phase $\Phi$, as in the simple case already discussed in
\ref{spontaneous}.\ The information on the phase is therefore a combination of
the information contained in the initial state and of that created by the
successive position measurements.

A classical example is a \textquotedblleft coherent state\textquotedblright%
\ obtained when the coefficients $x_{Q}$ are given by:%
\begin{equation}
x_{Q}=e^{-\left\vert \alpha\right\vert ^{2}/2}~\frac{\alpha^{Q}}{\sqrt{q!}}
\label{26}%
\end{equation}
where $\alpha$ is a parameter, which can be written:%
\begin{equation}
\alpha=\left\vert \alpha\right\vert ~e^{i\Theta} \label{27}%
\end{equation}
The initial $\Phi$ distribution is then given by:%
\begin{equation}
G(\Phi)\sim\left\vert \sum_{Q}\frac{\left\vert \alpha\right\vert ^{Q}}%
{\sqrt{q!}}~e^{iQ(\Phi-\Theta)}\right\vert ^{2} \label{28}%
\end{equation}
which is a narrow distribution around $\Phi=\Theta$ when $\left\vert
\alpha\right\vert ~\gg1$\ ; in this case, the initial value of $\Phi$ is
already very well defined, so that not much additional information can be
gained in successive interference measurements.

\subsection{General state}

In the most general case, both the sum and difference of $N_{a}$ and $N_{b}$
vary in the initial state:%
\begin{equation}
\mid\Phi_{0}>~=%
%TCIMACRO{\dsum \limits_{N}}%
%BeginExpansion
{\displaystyle\sum\limits_{N}}
%EndExpansion
~%
%TCIMACRO{\dsum \limits_{Q}}%
%BeginExpansion
{\displaystyle\sum\limits_{Q}}
%EndExpansion
x_{N_{a}~,~\ N_{b}}\mid N_{a}:\mathbf{k}_{a};~N_{b}:\mathbf{k}_{b}~>
\label{29}%
\end{equation}
where, under the sum, $N_{a}$ and $N_{b}$ are given by:%
\begin{equation}
N_{a}=\frac{N}{2}+Q\text{ \ \ \ ; \ \ \ }N_{b}=\frac{N}{2}-Q \label{30}%
\end{equation}
In the calculation of the probability, the variables $N$ and $Q$ play a
different role, because the sequence of $a$ and $a\dagger$ operators acts on
$Q$ but not on $N$.\ As a consequence, a double sum over $Q$ and $Q^{^{\prime
}}$ still appears, as in the preceding section, but only a single sum over
$N$.\ For each value of $N$, the calculation is the same as above, and leads
to the definition of a $N$ dependent distribution function for $\Phi$:%
\begin{equation}
G_{N}(\Phi)=\left\vert \sum_{Q}x_{(N/2)+Q,~(N/2)-Q}~~e^{-iQ\Phi}\right\vert
^{2} \label{31}%
\end{equation}
Then the sum over $N$ can be made in a second step; one can easily see that
equation (\ref{25}) remains valid provided one introduces the following
definition of $G(\Phi)$:%
\begin{equation}
G\left(  \Phi\right)  =\sum_{N}~G_{N}(\Phi) \label{32}%
\end{equation}
We therefore find a situation where each value of $N$ contributes
independently to the distribution of the phase $\Phi$. Different situations
are possible.\ If for instance all functions $G_{N}(\Phi)$ are peaked around
the same value of the phase, we obtain a state where the initial phase is well
defined, as for the coherent state (\ref{26}).\ But if, on the contrary, they
are either non-peaked functions, or functions peaked around different values,
the initial phase is uncertain; it only becomes better and better known as
more and more position measurements are performed, as in \ \S \ref{simple}.
This shows that the spontaneous appearance of relative phase that we have
discussed is not restricted to Fock states; it is actually a general property
of all states involving high population numbers, provided the number of
measurements remains much smaller than these populations.

\section{Spin states; quantum non-locality}

\label{spin}

We now apply our calculation to particles with two internal states, which we
note $\alpha$ and $\beta$.\ We consider that these states are spin $1/2$
states; the notion of pseudo spin allows us to do so without loss of generality.

\subsection{Calculation of the probability}

\label{probability}

The initial state of the system is assumed to be:%
\begin{equation}
\mid\Phi_{0}>~=~\mid N_{a}:\mathbf{k}_{a},\alpha~;~N_{b}:\mathbf{k}_{b}%
,\beta~> \label{33}%
\end{equation}
We note $\Psi_{\mu}(\mathbf{r})$, with $\mu=\alpha,\beta$, the field operators
associated with internal states $\alpha$,$\beta$.\ The $\mathbf{r}$ dependent
local density is then:%
\begin{equation}
n(\mathbf{r})=~~\Psi_{\beta}^{\dagger}(\mathbf{r})\Psi_{\beta}(\mathbf{r}%
)+\Psi_{\alpha}^{\dagger}(\mathbf{r})\Psi_{\alpha}(\mathbf{r}) \label{34}%
\end{equation}
and the three components of the local spin density are:%
\begin{equation}%
\begin{array}
[c]{l}%
\sigma_{z}(\mathbf{r})=~\Psi_{\beta}^{\dagger}(\mathbf{r})\Psi_{\beta
}(\mathbf{r})-\Psi_{\alpha}^{\dagger}(\mathbf{r})\Psi_{\alpha}(\mathbf{r})\\
\sigma_{x}(\mathbf{r})=~\Psi_{\beta}^{\dagger}(\mathbf{r})\Psi_{\alpha
}(\mathbf{r})+\Psi_{\alpha}^{\dagger}(\mathbf{r})\Psi_{\beta}(\mathbf{r})\\
\sigma_{y}(\mathbf{r})=~i\left[  \Psi_{\beta}^{\dagger}(\mathbf{r}%
)\Psi_{\alpha}(\mathbf{r})-\Psi_{\alpha}^{\dagger}(\mathbf{r})\Psi_{\beta
}(\mathbf{r})\right]
\end{array}
\label{35}%
\end{equation}
The spin component in the direction of plane $xOy$ making an angle $\theta$
with $Ox$ is:%
\begin{equation}
\sigma_{\theta}(\mathbf{r})=~e^{i\theta}\Psi_{\beta}^{\dagger}(\mathbf{r}%
)\Psi_{\alpha}(\mathbf{r})+e^{-i\theta}\Psi_{\alpha}^{\dagger}(\mathbf{r}%
)\Psi_{\beta}(\mathbf{r}) \label{36}%
\end{equation}
This operator, in spin space, has two eigenvalues $\eta=\pm1$; the
corresponding projectors are:%
\begin{equation}
P_{\eta}(\mathbf{r,\theta})=\,\frac{1}{2}\left[  n(\mathbf{r})+\eta
\sigma_{\theta}(\mathbf{r})\right]  \label{37}%
\end{equation}

We now consider a measurement of one spin at point $\mathbf{r}_{1}$ in
direction $\theta_{1}$, another at point $\mathbf{r}_{2}$ in direction
$\theta_{2}$, etc.; this corresponds to the sequence of operators:%
\begin{equation}
\sigma_{\theta_{1}}(\mathbf{r}_{1})\text{ \ ; \ }~\sigma_{\theta_{2}%
}(\mathbf{r}_{2})\text{ \ ; \ }~\sigma_{\theta_{3}}(\mathbf{r}_{3}%
)...~\sigma_{\theta_{P}}(\mathbf{r}_{P}) \label{38}%
\end{equation}
As in \ref{spontaneous}, we assume assume that all $\mathbf{r}$'s are
different so that all these operators commute; in addition, as already
mentioned, in all this article we assume that the sequence of measurements is
sufficiently brief to ignore any intrinsic evolution of the
condensates.\ Using again the expansion of the field operators on the
annihilation operators:%
\begin{equation}
\Psi_{\mu}(\mathbf{r})~\sim~\sum_{\mathbf{k}}e^{i\mathbf{k}\cdot\mathbf{r}%
}~a_{\mathbf{k;~\mu}} \label{39}%
\end{equation}
we obtain the probability of a sequence of results:%
\begin{equation}
\eta_{1}=\pm1\text{\ \ ; \ }~\eta_{2}=\pm1\text{ \ ; \ }...~\eta_{P}=\pm1
\end{equation}
in the form of an average value in state $\mid\Phi_{0}>$:%
\begin{equation}%
\begin{array}
[c]{l}%
<\Phi_{0}\mid P_{\eta_{1}}(\mathbf{r}_{1},\theta_{1})\times P_{\eta_{2}%
}(\mathbf{r}_{2},\theta_{2})\times.....\times P_{\eta_{P}}(\mathbf{r}%
_{P},\theta_{P})\mid\Phi_{0}>\\
\sim\sum_{\mathbf{k}_{1}}\sum_{\mathbf{k}_{1}^{^{\prime}}}~e^{i(\mathbf{k}%
_{1}-\mathbf{k}_{1}^{^{\prime}})\cdot\mathbf{r}_{1}}\sum_{\mathbf{k}_{2}}%
~\sum_{\mathbf{k}_{2}^{^{\prime}}}e^{i(\mathbf{k}_{2}-\mathbf{k}_{2}%
^{^{\prime}})\cdot\mathbf{r}_{2}}\times..\\
<\Phi_{0}\mid%
%TCIMACRO{\dprod \limits_{i}}%
%BeginExpansion
{\displaystyle\prod\limits_{i}}
%EndExpansion
\left[  a_{\mathbf{k}_{i}^{^{\prime}};\beta}^{\dagger}a_{\mathbf{k}_{i}^{{}%
};\beta}^{{}}+a_{\mathbf{k}_{i}^{^{\prime}};\alpha}^{\dagger}a_{\mathbf{k}%
_{i};\alpha}^{{}}+\eta_{i}\left(  ~e^{i\theta_{i}}~a_{\mathbf{k}_{i}%
^{^{\prime}};\beta}^{\dagger}a_{\mathbf{k}_{i}^{{}};\alpha}^{{}}%
+~e^{-i\theta_{i}}~a_{\mathbf{k}_{i}^{^{\prime}};\alpha}^{\dagger
}a_{\mathbf{k}_{i}^{{}};\beta}^{{}}\right)  \right]  \mid\Phi_{0}>
\end{array}
\label{41}%
\end{equation}

The rest of the calculation is now very similar to that of \S \ref{simple}%
.\ In each term contained in the product of the second line of (\ref{41}), we
may assume that the $a$ and $a^{\dagger}$ are ordered in the \textquotedblleft
normal\textquotedblright\ order (with all the $a$'s operators to the right,
all $a^{\dag}$'s to the left). This is because, in the first line of
(\ref{41}), when the projectors are replaced by their expressions in function
of the field operators (and hermitian conjugate), one can move all $\Psi$'s to
the right, all $\Psi^{\dag}$'s to the left (since operators at different
points of space commute).\ Now we see that, each time a $\mathbf{k}_{i}$ or a
$\mathbf{k}_{i}^{^{\prime}}$ appears associated with the spin index $\alpha$
(or $\beta$), it is necessarily equal to $\mathbf{k}_{a}$ (or $\mathbf{k}_{b}%
$) to give an non-zero contribution; the operators $a$ and $a^{\dagger}$
introduce numbers $\sqrt{N_{a,b}\pm n}$ as before, the only condition being
that the creation and annihilation operators should be balanced in each
sequence.\ If we assume as above that the number of measurements $P$ is much
smaller than $N$, the second line becomes proportional to:%
\begin{equation}
\sim\int_{0}^{2\pi}\frac{d\Phi}{2\pi}~%
%TCIMACRO{\dprod \limits_{i=1}^{P}}%
%BeginExpansion
{\displaystyle\prod\limits_{i=1}^{P}}
%EndExpansion
\left[  1+\frac{x}{2}~\eta_{i}~~e^{i\left[  (\mathbf{k}_{a}-\mathbf{k}%
_{b})\cdot\mathbf{r+}\theta_{i}\right]  }~e^{-i\Phi}+c.c.)\right]  \label{42}%
\end{equation}
where the contrast ratio $x$ is defined by equation (\ref{9}). The probability
of the sequence of results $\eta_{1}$, $\eta_{2}$, ..$\eta_{P}$ obtained at
points $\mathbf{r}_{1}$, $\mathbf{r}_{2}$ , ..$\mathbf{r}_{P}$ is then:\ %

\begin{equation}
\sim\int_{0}^{2\pi}\frac{d\Phi}{2\pi}~%
%TCIMACRO{\dprod \limits_{i=1}^{P}}%
%BeginExpansion
{\displaystyle\prod\limits_{i=1}^{P}}
%EndExpansion
\left\{  1+x~\eta_{i}\cos\left[  (\mathbf{k}_{a}-\mathbf{k}_{b})\cdot
\mathbf{r}_{i}+\theta_{i}-\Phi\right]  \right\}  \label{43}%
\end{equation}
which is the general result.

From now on in this section, we assume that $N_{a}=N_{b}$, so that the
contrast ratio $x$ is equal to $1$. For a sequence of $+1$ results, (\ref{43})
reduces to:%
\begin{equation}
\sim\int_{0}^{2\pi}\frac{d\Phi}{2\pi}~%
%TCIMACRO{\dprod \limits_{i=1}^{P}}%
%BeginExpansion
{\displaystyle\prod\limits_{i=1}^{P}}
%EndExpansion
\cos^{2}\left[  \frac{(\mathbf{k}_{a}-\mathbf{k}_{b})\cdot\mathbf{r_{i}%
+}\theta_{i}-\Phi}{2}\right]  \label{44}%
\end{equation}

A simpler case occurs when $\mathbf{k}_{a}=\mathbf{k}_{b}$.\ The probability
of a sequence of $P_{+}$ results equal to $+1$ and $P_{-}$ results equal to
$-1$ then becomes:%
\begin{equation}
\sim\int_{0}^{2\pi}\frac{d\Phi}{2\pi}%
%TCIMACRO{\dprod \limits_{1}^{P_{+}}}%
%BeginExpansion
{\displaystyle\prod\limits_{1}^{P_{+}}}
%EndExpansion
\cos^{2}(\frac{\theta_{i}-\Phi}{2})%
%TCIMACRO{\dprod \limits_{1}^{P_{-}}}%
%BeginExpansion
{\displaystyle\prod\limits_{1}^{P_{-}}}
%EndExpansion
\sin^{2}(\frac{\theta_{i}-\Phi}{2}) \label{45}%
\end{equation}
or, if all angles $\theta$ are equal:%
\begin{equation}
\sim\int_{0}^{2\pi}\frac{d\Phi}{2\pi}~\left[  \cos\frac{\Phi}{2}\right]
^{2P_{+}}~\left[  \sin\frac{\Phi}{2}\right]  ^{2P_{-}} \label{46}%
\end{equation}
(Wallis integral).

If we compare with the situation studied in \S \ref{spontaneous}, we now have
an additional element: the adjustable parameter $\theta_{i}$ (direction of the
measured spin component in the $xOy$ plane).\ This combines with the
$\mathbf{r}_{i}$ dependent interference effect to determine the probability of
each sequence of results, as shown by equation (\ref{43}).\ When
$\mathbf{k}_{a}=\mathbf{k}_{b}$, these spatial interference effects disappear,
and the probability of the succession of results takes the simpler form
(\ref{45}), where the effect of the $\eta$'s and $\theta$'s is isolated. For
the first measurement, the two results $\eta=\pm1$ are equally probable but,
as soon as one of these results is obtained, the distribution of $\Phi$ is
changed.\ The basic process behind the progressive determination of $\Phi$ is
the same as in \S \ref{spontaneous}: after a given sequence of measurements
with known results, the new distribution of probability for $\Phi$ is given
(within a normalization constant) by the initial probability of this sequence,
i.e. by replacing in (\ref{43}) the measured $\eta_{i}$'s by the corresponding
results. Each result $\eta_{i}=+1$ brings in a factor $\cos^{2}(\theta
_{i}-\Phi)/2$ that tends to localize the $\Phi$ distribution around
$\Phi=\theta_{i}$ and makes it vanish in the opposite direction $\Phi
=\theta_{i}+\pi$; each result $\eta_{i}=-1$ brings in a factor $\sin
^{2}(\theta_{i}-\Phi)/2$ that does exactly the opposite. A combination of two
successive opposite results $\eta=+1,-1$ brings in a factor $\sim\sin
^{2}(\theta_{i}-\Phi)$ that cancels the $\Phi$ distribution in the two
directions $\Phi=\theta_{i}$ and $\Phi=\theta_{i}+\pi$ and tends to localize
it around $\Phi=\theta_{i}\pm\pi/2$. Similarly, sequences of $\eta_{i}$s with
different numbers of results $+1,-1$ will introduce a $\Phi$ maximum in some
intermediate direction (see for instance the discussion of \S III-C of ref.
\cite{CD}).

This illustrates how a sequence of successive measurements can progressively
make the $\Phi$ distribution more and more peaked around some value $\Phi
=\Phi_{\max}$, and brings the system closer to a situation where all the spins
are polarized in the same transverse direction $\theta=\Phi_{\max}$.;
interesting numerical simulations of the phenomenon can be found in reference
\cite{JJ}. The individual discrete measurements cooperate to provide
information on a quasi-continuous quantity, the direction of the transverse
orientation.\ Moreover, the experimenter can adjust the choice of the
measurement angles $\theta_{i}$ in function of the preceding results $\pm1$
obtained.\ It is not necessary to orient the measurement apparatus in the
direction of the component that is to be measured, as would be the case for a
single spin; a possible strategy to measure the position of the maximum
$\Phi_{\max}$ is to perform measurements in the perpendicular direction and
check that the proportion of $\ \eta\pm1$ results is $1/2$ (in this way, first
order $\theta$ variations can be measured).

\subsection{Quantum amplification}

\label{ampli}

An interesting property of the spin measurement sequence is that it can give
information on the transverse spin orientation of a very large number $2N$ of
particles even when $P$, the number of actually measured particles, is much
lower.\ In principle, one can get information on the spin orientation of a
macroscopic sample by just measuring the spin orientation of a microscopic
sub-sample made of $100$ particles for instance, which involves a huge
amplification factor.\ Such an amplification is often evoked in the
theoretical discussion of the measurement process in quantum mechanics, where
a microscopic system (the measured system) triggers an instability in the
measurement apparatus, leading to macroscopically different states of the
pointer (here the large number of non-measured particles).\ The phenomenon is
clearly related to the effect discussed by Siggia and Ruckenstein \cite{SR} in
the context of the spontaneous appearance of a transverse spin orientation in
a double condensates of spin polarized atomic hydrogen.

A related discussion is given by Leggett and Sols in ref. \cite{LS} (see
paragraph before last of this reference), who consider two superconducting
systems that are initially described by an incoherent mixture of number
states, and between which a Josephson current can flow.\ They then ask the
question \textquotedblleft does the act of looking to see whether a current
flows itself force the system into an eigenstate of current and hence of
relative phase?\textquotedblright.\ They assume the presence of a small
compass needle measuring the magnetic field produced by the Josephson
current.\ Both the needle and the current are macroscopic, but one can assume
that the current is arbitrarily large while the needle is tiny.\ Under these
conditions, it seems \textquotedblleft bizarre in the
extreme\textquotedblright\ to assume that, by some mysterious amplification
effect, it is the small classical object that will force the large one to take
a definite value.\ The authors conclude that, while standard mechanics would
answer \textquotedblleft yes\textquotedblright\ to their question,
\textquotedblleft common sense rebels against this conclusion, ... and we
believe that in this case common sense is right\textquotedblright\ - in other
words they conclude that the phase $\Phi$ existed before the measurement
(additional variable).\ In our case, the discussion is different since no
external classical pointer (the compass needle) is introduced.\ In a sense,
the point is even stronger since the small object forcing the large classical
object into a definite value may even be microscopic.\ But the conclusion
remains the same: if one rejects the idea of a small object creating the value
of a macroscopic variable (the phase), then one is led naturally to accept the
existence of hidden variables in quantum theory. We note in passing that, as
remarked by Bell \cite{Bell-1, Bell-2}, even if it is traditional, it is
somewhat clumsy to call \textquotedblleft hidden\textquotedblright\ an
additional variable such as $\Phi$: it is precisely the variable that is
actually seen in the experiment, while the variables of quantum mechanics
(wave functions) tend to remain invisible.

\subsection{Non-local quantum effects}

For the discussion of this subsection, it is convenient to assume that the
orbital states of the particles are not necessarily plane waves.\ We therefore
replace (\ref{33}) by:%
\begin{equation}
\mid\Phi_{0}>~=~\mid N_{a}:\varphi_{a},\alpha;~N_{b}:\varphi_{b},\beta~>
\label{47}%
\end{equation}
where the orbital states correspond to the wave functions:%
\begin{equation}
<\mathbf{r}\mid\varphi_{a}>=\varphi_{a}(\mathbf{r})\text{ \ \ },\text{
\ \ \ }<\mathbf{r}\mid\varphi_{b}>=\varphi_{b}(\mathbf{r}) \label{48}%
\end{equation}
with relative phase given by:%
\begin{equation}
\arg\left\{  \varphi_{a}(\mathbf{r})/\varphi_{b}(\mathbf{r})\right\}
=\xi(\mathbf{r}) \label{49}%
\end{equation}

The calculation remains similar to that given above.\ In (\ref{39}), the field
operator was expanded on the annihilation operators $a_{\mathbf{k}}$
corresponding to plane waves; here we expand $\Psi_{\alpha}(\mathbf{r})$ on
annihilation operators corresponding to a base where the particle is in
internal state $\alpha$ and in orbital states with wave functions $\varphi
_{a}(\mathbf{r})$, $\varphi_{2}(\mathbf{r})$, $\varphi_{3}(\mathbf{r})$, etc.:%
\begin{equation}
\Psi_{\mu}(\mathbf{r})~\sim~\sum_{q\mathbf{=1}}^{\infty}\varphi_{q}%
(\mathbf{r})~a_{q\mathbf{;~\mu}} \label{49bis}%
\end{equation}
with$~~~\varphi_{1}(\mathbf{r})=\varphi_{a}(\mathbf{r})$, and where
$\varphi_{2}(\mathbf{r})$, $\varphi_{3}(\mathbf{r})$, etc. complete
$\varphi_{a}(\mathbf{r})$ in order to make an orthonormal basis in orbital
space.\ We do something similar for $\Psi_{\beta}(\mathbf{r})$, now in a basis
with orbital states that contain $\varphi_{b}(\mathbf{r})$ as the first
vector.\ In practice, the only difference is that the exponentials
$e^{i\left[  (\mathbf{k}_{a}-\mathbf{k}_{b})\cdot\mathbf{r}_{i}\right]  }$ are
now replaced by the product $\varphi_{b}^{\ast}(\mathbf{r})\varphi
_{a}(\mathbf{r})$.\ The result for the probability is then:%
\begin{equation}
\sim\int_{0}^{2\pi}\frac{d\Phi}{2\pi}~%
%TCIMACRO{\dprod \limits_{i=1}^{P}}%
%BeginExpansion
{\displaystyle\prod\limits_{i=1}^{P}}
%EndExpansion
\left[  \left\vert \varphi_{a}(\mathbf{r}_{i})\right\vert ^{2}+\left\vert
\varphi_{b}(\mathbf{r}_{i})\right\vert ^{2}+2x\eta_{i}\left\vert \varphi
_{a}(\mathbf{r}_{i})\varphi_{b}(\mathbf{r}_{i})\right\vert ~\cos\left[
\xi(\mathbf{r}_{i})-\theta_{i}+\Phi\right]  \right]  \label{50}%
\end{equation}
Even if $N_{a}=N_{b}$ (so that $x=1$) the contrast ratio of the interference
term, which contains the $\theta_{i}$ dependence of the result, is not
necessarily $1$; it is maximum at points $\mathbf{r}$ where they have the same
modulus, zero if one of them vanishes.\ Clearly, in order to determine the
relative phase $\Phi$, transverse spin measurements should be made in the
regions of good overlap between $\varphi_{a}(\mathbf{r})$ and $\varphi
_{b}(\mathbf{r})$,

We can now assume that $\varphi_{a}(\mathbf{r})$ and $\varphi_{b}(\mathbf{r})$
are equal and that both these functions are non-zero only in two distant
regions of space, $D$ and $D^{^{\prime}}$; for simplicity we will assume that
they are constant within these two domains; equation (\ref{50}) then
simplifies and the orbital wave functions disappear, as in (\ref{45}).\ All
the measurements are made in region $D$, which is supposed to be relatively
small so that it contains a small average number of particles, $10$ or $100$
for instance.\ Region $D^{^{\prime}}$ can be much larger and contain a
macroscopic number of particles. Both regions can be very distant from each
other, so that no signal transmission between them is possible during the
duration of the experiment. The curious prediction of quantum mechanics is
that a measurement of a microscopic transverse spin orientation in region $D$
will immediately induce the appearance of a parallel macroscopic spin
orientation in region $D^{^{\prime}}$.

One can make the situation even more striking by assuming that the wave
functions $\varphi_{a,b}(\mathbf{r})$ are non-zero in three regions of space:
a small measurement region $D$ as before, a large region $D^{^{\prime}}$ at a
small distance from $D$, and finally another large region $D^{^{\prime\prime}%
}$ very far away (in a different galaxy for instance).\ The transverse spin
orientation in $D^{^{\prime}}$ then plays the role of a local pointer, which
can be seen as a part of the measurement apparatus which measures the
direction of the transverse spin orientation in $D$. Remote region
$D^{^{\prime\prime}}$ contains another pointer, which under the effect of a
measurement in $D$ immediately takes a direction parallel to that of the
pointer in $D^{^{\prime}}$; the phenomenon could then be called teleportation
of the direction of pointers, a clear illustration of quantum
non-locality.\ Of course, we should also make the usual proviso: the direction
obtained in the measurement can not be controlled, but is completely random:
there is no way to use the teleportation to transmit information \cite{BE,
Eberhard}, a necessary condition for preserving consistency with relativity.

\subsection{The EPR argument for condensates}

The perfect correlation between the direction of pointers at arbitrary
distances is of course reminiscent of the quantum correlations discussed in
the context of the famous Einstein-Podolsky-Rosen (EPR) argument; the
similarity is even more striking in the EPRB\ version proposed by Bohm (B),
which involves the measurement of spin directions.\ For a review, see for
instance ref. \cite{FL-1} and references therein; a detailed historical
perspective is given in \cite{Jammer}. EPRB consider two correlated particles
which undergo spin measurements in two remote regions of space and assume that
quantum mechanics gives correct predictions concerning the probabilities of
the various measurements of the spins along different directions.\ The EPR
reasoning \cite{EPR} states that \textquotedblleft..if, without in any way
disturbing a system we can predict with certainty the value of a physical
quantity, then there exists an element of physical reality corresponding to
this quantity\textquotedblright\ (may the most quoted sentence of all
physics!).\ Since EPR attribute independent elements of physical reality to
the content of remote regions of space, this is often called an assumption of
\textquotedblleft local realism\textquotedblright.\ From this they show that
some elements of physical reality are not contained in quantum mechanics, in
other words that this theory is incomplete. Bohr rejected this conclusion
\cite{Bohr} because he considered that the notion of element of reality used
by EPR\ contained essential ambiguities.\ In his view, \textquotedblleft the
procedure of measurement has an essential influence on which the very
definition of the physical quantities in question rests\textquotedblright%
.\ One should then only ascribe physical reality to the whole extended system
made of the microscopic particles \textit{and} the macroscopic measurement
apparatus, not to microscopic subsystems only. As a consequence, one can not
always attribute distinct physical properties to the content of in different
regions of space, as EPR do.\ In other words, Bohr sees the process of
measurement on an extended microscopic system as a fundamentally random
process that is not necessarily localized in space \cite{FL-2}.

Let us now apply the EPR reasoning to the evolution of the macroscopic
pointers (transverse spin orientation) in $D^{\prime}$ and $D^{^{\prime\prime
}}$, when a series of spin measurements is performed in $D$. We focus for a
moment on $D^{^{\prime\prime}}$ and consider the elements of physical reality
associated with this region of space.\ As EPR, we consider that they can not
vary suddenly under the effect of events taking place at arbitrarily large
distances in $D$ and $D^{^{\prime}}$.\ Now compare the elements of reality
associated to region $D^{^{\prime\prime}}$, just before, and just after the
series of spin measurements performed in region $D$. After the measurement the
region contains a macroscopic spin orientation in a given transverse
direction, which are necessarily associated with some elements of
reality.\ But these elements cannot appear as a result of a random process
taking place in $D$, with no possible causal link since the distance is much
too large. Therefore, even before the measurement, the element(s) of reality
associated with the spin orientation already existed.\ But, on the other hand,
standard quantum mechanics does not include this element before the
measurement, only after; it therefore misses at least one element of reality,
it is incomplete.

One can make the reasoning more explicit by considering two realizations of
the experiment, one in which the transverse spin orientation is found at the
end of the experiment in a direction $\Phi=\Phi_{1}$ (with some small
uncertainty $\Delta\Phi$ depending on the number or measurements), another in
which this spin orientation is found in a different direction $\Phi=\Phi_{2}$,
after exactly the same sequence of measurements. Standard quantum mechanics
describes the two final situations by different vectors in the space of states
of the system, accounting for the existence of different physical quantities
in region $D^{^{\prime\prime}}$(different elements of reality).\ On the other
hand, it considers that the initial states were exactly the same; the
differences appear only under the effect of the measurements. But this is in
opposition with the EPR\ notion of local reality, which implies that the
differences must have already existed before the measurements.\ In other
words, the EPR reasoning shows that the two realizations of the experiment
actually started from different initial conditions.\ Since quantum mechanics
ignores this difference, it is incomplete.

What are the consequences of this transposition of the EPR\ reasoning to a
different situation than the two particles of the original article?\ First,
the EPR sentence quoted in the first paragraph of this subsection becomes even
more true.\ Here the pointers in $D^{^{\prime}}$ and $D^{^{\prime\prime}}$ are
not invisible microscopic objects, but macroscopic and permanent angular
momenta, similar to the spin magnetization of ferromagnets; they are classical
physical entities that can be manipulated directly, so that it seems difficult
to deprive them of properties that are independent of the measurement
apparatuses.\ Therefore Bohr's denial of independent physical reality becomes
more difficult to accept (of course, no one knows if Bohr would have given
again it in this context!).\ Second, even if one follows Bohr and takes for
granted that physical properties are reserved to systems including macroscopic
measurement apparatuses, another difference arises here: one single
experimental setup can lead to many spin orientations.\ With a single spin,
one can measure only the spin component along the direction defined by the
apparatus; with many spins in region $D$, it is the measurement process which
determines the axis along which the spins become oriented (the transverse
orientation can take in general a direction which makes any angle with the
direction of measurement).\ In order to show that incompatible observables can
simultaneously be defined (as opposed to standard quantum mechanics), the
usual EPR argument involves different (and incompatible) experimental
setups.\ Here, no measurement at all is performed in $D^{^{\prime}}$ and
$D^{^{\prime\prime}}$; moreover, the sequence of measurements performed in $D$
may be always exactly the same in successive realizations of the experiment,
even if they lead to different directions of the transverse orientation
\footnote{If region $D^{^{\prime}}$ contains a large average number or
particles, its spin orientation is classical; within a good approximation, all
its components correspond to communing operators, and they are not
incompatible quantities in standard quantum mechanics.\ On the other hand, if
regions $D^{^{\prime}}$ contains only one or two particles on average,
orthogonal components of the spin orientation are no longer compatible
observables, allowing one to transpose the usual EPR argument on incompatible
experiments to this case.}. Clearly, Bohr's argument ascribing different
physical realities to incompatible measurements does not apply anymore when
only one kind of measurement takes place.\ \ Another difference is that, in
the usual discussion of EPR correlations, the entanglement of the two spin
particles plays an essential role; here, the initial state vector is a simple
product involving the two internal spin states.\ One can finally remark that,
in usual EPRB situations, one has to invoke a \textquotedblleft no quantum
cloning\textquotedblright\ theorem to explain why superluminal signalling is
impossible.\ Here, in all regions $D$, $D^{^{\prime}}$ and $D^{^{\prime\prime
}}$, we have a large number of copies of the individual systems to measure
precisely the transverse direction of polarization.

The EPR argument therefore certainly looks different, and probably even
stronger, in the context of Bose-Einstein spin condensates.\ Within standard
quantum mechanics, it seems really difficult not to speak of action at a
distance to describe the instantaneous appearance of the macroscopic
transverse orientation of the pointer in region $D^{^{\prime\prime}}$ - but,
again, we must remember that its final direction is totally uncontrolled and
cannot be used as a signal.\ In other words, what we we have is an
\textquotedblleft uncontrolled action at a distance\textquotedblright.\ On the
other hand, we do not have the equivalent of a Bell theorem in this case: the
introduction of an additional variable $\Phi$ to quantum mechanics easily
allows one the reproduce the prediction of quantum mechanics in a purely local
model. Amusingly, the situation is exactly the opposite of that usually
discussed in the context of EPR-Bell experiments on pairs of spins: here
non-locality appears clearly in standard quantum mechanics, but disappears in
theories with additional variables.

\subsection{Conservation of angular momentum}

An intriguing question relates to the momentum absorbed by the measurement
apparatus during the interaction with the measured system.\ \ For instance, in
a usual Stern-Gerlach experiment where a single spin, initially polarized in a
transverse direction, is measured to end up in a longitudinal direction (with
respect to the measurement apparatus), it is generally assumed that the
variation of its angular momentum is absorbed by the apparatus - this
variation is microscopic and so small that it is impossible to measure in
practice.\ Here, we have a system which starts before measurement from a state
(\ref{47}) with zero average value of the angular momentum \footnote{We reason
within standard quantum mechanics, assuming that it is the measurement process
that puts the system into a state with definite angular momentum; we do not
assume that the process only reveals a value that existed before, as in
theories with additional variables.}; after a sequence of measurements
performed in $D$ it reaches another state where all spins point in a
transverse direction given by the spontaneous value of $\Phi$, which has a
high angular momentum if there are many spins. From the usual conservation
rule, one could then expect a priori that the measurement apparatus should
acquire the opposite amount of angular momentum.\ On the other hand, we have
seen that the measurement of a few spins in $D$ may result in the appearance
of a macroscopic angular momentum for the whole system in $D+D^{^{\prime}}%
$.\ Does the measurement apparatus really absorbs all the angular momentum
acquired by the whole system, although it interacts directly only with a
subsystem in $D$ that is very small?

The usual rules of quantum mechanics are not very specific about the answer to
this question. The emphasis is usually put on measurements of microscopic
systems, which have a completely negligible effect on the macroscopic
measurement apparatus - an exception, nevertheless, is the discussion of the
recoil effect of a moving double slit experiment in the famous Bohr-Einstein
debate.\ Here, due to the amplification process discussed in \S \ref{ampli},
the measured system itself is macroscopic.\ If we assume that the answer to
the question is \textquotedblleft yes\textquotedblright,\ in order to keep a
strict conservation rule for the angular momentum, we open the way to
paradoxes: for instance, by locally dephasing in region $D^{^{\prime}}$ the
two wave functions $\varphi_{a}$ and $\varphi_{b}$, before the measurement in
$D$, one would control the direction of the angular momentum in $D^{^{\prime}%
}$ induced by this measurement, and therefore the transfer of angular momentum
to the apparatus in region $D$. This would allow superluminal signalling, in
contradiction with relativity.\ The appropriate answer is therefore probably
rather \textquotedblleft no, the angular momentum transferred is just that of
the individual spins that are measured\textquotedblright; in this perspective,
the measurement operation creates in region $D^{^{\prime}}$ an angular
momentum that appears from nothing, which is somewhat paradoxical too.\ Maybe
this just means that the rule of angular momentum conservation should to not
be taken too strictly in a quantum measurement processes. Another possibility,
again, is to take the point of view of theories with additional variables, and
assume that $\Phi$ was initially actually perfectly determined in each
realization of the experiment, even if it is unknown; in this perspective no
angular momentum transfer has to take place, so that the difficulty vanishes.

\subsection{More than two internal levels}

The above considerations can easily be transposed to more than two internal
states, three for instance.\ The situation then becomes analogous to that of
\S \ref{three}, where two relative phases $\Phi$ and $\Phi^{^{\prime}}$ enter
the calculation instead of one.\ One can assume that transverse spin
observables relative to the $\alpha-\beta$, and to the $\beta-\gamma$,
transitions are measured.\ A sufficient number of measurements will determine
the two relative phases.\ An interesting property is that the transverse
orientation of the spin associated to the $\alpha-\gamma$ transition will then
be determined, although no direct measurement of this observable has been made.

\section{Conclusion}

In standard quantum mechanics, physical systems in highly populated Fock
states have a completely undetermined relative phase, but this phase can
become very well defined under the effect of a few interference measurements
only; this is true even in the number of particles detected is negligible with
respect to the total populations.\ The results mimic exactly what would happen
if the initial phase was fixed, but initially unknown, and progressively
revealed by the observation of the interference pattern.\ So, the answer to
the question mentioned in the introduction \textquotedblleft do condensates
that have never seen each other have a relative phase?" is: in standard
quantum mechanics, in principle they do not, but they can acquire it under the
effect of a few microscopic measurements, in a way which gives the impression
of a well defined initial phase.\ As Anderson puts it \cite{A-3}:
\textquotedblleft any future experiment can be interpreted as if $\Phi$ was
fixed\textquotedblright.\ It is therefore tempting to assume that the phase
existed from the beginning.\ This is the point of view of physicists who argue
that, for superfluids, the usual postulates of quantum mechanics should be
completed by an additional postulate introducing spontaneous symmetry breaking
(of the conservation of the number of particles); this leads to non-zero
average value of the field operator.\ This view is of course legitimate, but
is actually nothing but another form of the \textquotedblleft hidden
variable\textquotedblright\ theory of quantum mechanics (de Broglie, Bohm), as
we have discussed in \S \ \ref{simple}. It also remains legitimate to stick to
orthodox quantum mechanics and to consider that it is the process of quantum
measurement itself that acts on the condensates and forces them to acquire a
well defined relative phase.

Spin adds to this problem the notion of angular momentum and offers
interesting variants of the usual EPR non-locality experiments.\ One can
suppose that spin orientation measurements are performed in two (or more) very
remote regions of space.\ An interesting possibility is that some of the
remote systems may be macroscopic; the results in terms of transverse
direction of the local spin orientation of the sample are not discrete, as in
usual EPR\ experiments, but continuous.\ Actually the macroscopic spin
orientation plays here the role of the usual pointer often considered in the
theory of quantum measurement, but with a surprising property: the pointer can
be put at an arbitrary distance from the measured system.\ In one region of
space, a series of measurements determines a transverse orientation, in
another remote region a classical macroscopic pointer immediately takes the
corresponding direction and stays there, accessible to future
measurements!\ It seems difficult to deny that standard quantum mechanics
involves action at a distance in this case.\ This action is instantaneous, but
not controllable, so that it can not be used to send superluminal
signals.\ There has been a long debate \cite{Stapp-1, Mermin, Espagnat,
Shimony} to decide whether or not non-locality is an inherent property of
quantum mechanics, or if it appears only if additional ingredients are added
to the theory (hidden variables, local reality, countrafactuality
\cite{Stapp-2}, etc.). Our example can be discussed just in terms of standard
quantum mechanics, without adding such additional elements and/or incompatible
experimental setups, which would allow to invoke different physical realities
in a Bohr type argument.\ It therefore clearly speaks in favor of accepting
non-locality as an intrinsic property of standard quantum mechanics.

\bigskip

\begin{center}
Acknowledgments
\end{center}

This work was initiated during workshop \textquotedblleft BEC
2002\textquotedblright\ at the European Center for theoretical physics (ECT)
in Trento (Italy); the author is also grateful to Jean Dalibard, Claude
Cohen-Tannoudji and Jean-No\"{e}l Fuchs for useful discussions.

\end{document}